# Structural Phases of Bounded Three-Dimensional Screened Coulomb Clusters (Finite Yukawa System)

Ke Qiao, Matthew Benesh, and Truell Hyde, *Member, IEEE*

*Abstract*— The formation of three-dimensional (3D) dust clusters within a complex plasma modeled as a spatially confined Yukawa system is simulated using the box_tree code. Similar to unscreened Coulomb clusters, the occurrence of concentric shells with characteristic occupation numbers was observed. Both the occupation numbers and radii were found to depend on the Debye length. Ground and low energy meta-stable states of the shielded 3D Coulomb clusters were determined for 4<N<20. The structure and energy of the clusters in different states was analyzed for various Debye lengths. Structural phase transitions, including inter-shell structural phase transitions and intra-shell structural phase transitions, were observed for varying Debye length and the critical value for transitions calculated.

*Index Terms*—Complex plasma, coulomb cluster, structural phase transition, plasma crystal

## I. Introduction

Spatially confined charged particle systems include one component plasmas (OCP), e.g., electrons or ions in Paul and Penning traps [1]-[3], and complex (dusty) plasmas in discharge chambers [4]-[10]. Crystallization of such spatially confined charged particles has been observed in both OCP [1]-[3] and complex plasmas [4]-[10]. Since the interaction potential between grains is assumed to be either a Coulomb (OCP) or screened Coulomb (complex plasma) potential, such systems are generally referred to as Coulomb or screened Coulomb (Yukawa) systems. Crystalline systems consisting of a smaller number of particles (N ≤ 1000) are then called coulomb clusters and can form either two-dimensional (2D) [7], [8] or three-dimensional (3D) systems [2], [3], [9], [10], depending on the experimental environment. A large amount of research has recently been devoted into investigating the structure and dynamics of 2D coulomb clusters, both experimentally [7] and theoretically [8].

3D coulomb clusters, are much more difficult to examine, although experimental research has been conducted on ultra-cold ion plasmas confined in Paul and Penning traps [2], [3] and more recently, spherical dust coulomb clusters have been created in complex plasmas [9], [10]. Theoretical investigations into the structure and ground state energies of 3D coulomb clusters has also been conducted [11]-[13], but only for pure (unscreened) coulomb systems. Although this approach is suitable for a confined OCP such as that exhibited by a trapped ion plasma, it is well-known that the interaction potential between dust particles in complex plasmas is not a pure coulomb interaction. As shown in [10], a Coulomb ball in complex plasmas must instead be modeled as a Yukawa system, with a Yukawa potential

$$v(r) = q \exp(-r/\lambda_D)/4\pi\varepsilon r , \qquad (1)$$

assumed to act between dust particles. In Eq. (1), $q$ is the dust particle charge, $r$ is the distance between any two particles and $\lambda_D$ is the dust Debye length.

Structures of unbounded Yukawa systems have been investigated thoroughly and a phase diagram established (e.g., [14]) with these research results also compared to experimental research (e.g., [15]). However, since all Yukawa systems investigated experimentally are necessarily confined via an external potential, a theoretical investigation into the structural phase transitions of a bounded Yukawa system is essential for a proper understanding of the underlying physics involved. In addition, a study of spherical crystals interacting through Yukawa potential opens up an interesting new field which in a natural way bridges the gap between the above mentioned theoretical investigation into finite size Coulomb systems and the theory of macroscopic Yukawa plasmas, e.g. [10].

In this research, the crystallization of a 3D Yukawa system assuming a parabolic spatial confinement was simulated using the box_tree code [16]-[20]. Simulations were conducted for various screening characterized by the Debye length and the resulting cluster structure, energy and dependence on screening length analyzed.

The outline of the paper is as follows. In Sec, II, the box_tree code is briefly reviewed and the conditions under which simulations were conducted are explained. In Sec III, simulation results for 4 ≤ N ≤ 1400 are given and the

Manuscript received August 4, 2006. This work was supported in part by National Science Foundation Grant PHY-0353558.
Ke Qiao and Truell Hyde are with the Center for Astrophysics, Space Physics, and Engineering Research, Baylor University, Waco, TX 76798 USA (phone: 254-710-3763; e-mail: Ke_Qiao@baylor.edu, Truell_Hyde@baylor.edu).
Matthew Benesh was a 2006 REU fellow with CASPER, Baylor University, Waco, TX 76798 USA.

dependence of the cluster's structure and occupation number on the Debye length analyzed. In Sec. IV, the ground states and low energy meta-stable states are obtained from simulations for particle number $4 \leq N \leq 20$ and the energy dependence on Debye length analyzed. In Sec. V, the cluster structure for low energy states is analyzed and the inter-shell and intra-shell structural phase transitions as related to changing Debye length investigated. Conclusions are given in Sec. VI.

## II. NUMERICAL SIMULATION

The formation of 3D shielded Coulomb clusters was simulated for particle numbers ranging from $N = 4$ to $N = 1400$ using the box_tree code [16]-[20]. The box_tree code is a Barnes-Hut tree code written by Richardson [16] and later modified by Matthews and Hyde [17], Vasut and Hyde [18], and Qiao and Hyde [19], [20] to simulate complex plasmas under various conditions. Box_tree models systems composed of large numbers of particles by dividing the 3D box containing them into self-similar nested sub-boxes and then calculating all interparticle interactions using an incorporated tree code. Since the majority of these interparticle interactions can be determined by simply examining the multipole expansion of the collections of particles within the sub-boxes, the code scales as NlogN instead of $N^2$, resulting in much greater CPU time efficiency than is possible employing a traditional molecular dynamic approach.

In this research, the interparticle potential is modeled as a Yukawa potential of the form given by Eq. (1). In accordance with recent experiments on Coulomb balls [9], [10] and previous experiments and simulations on ion crystals [3], the external confining potential is assumed to be parabolic in 3D,

$$E_{ext}(x,y,z) = \frac{m}{2}\left[\omega^2(x^2 + y^2 + z^2)\right] \quad (2)$$

where $x, y, z$ are representative particle coordinates and $\omega^2$ is the measure of the strength of the parabolic confinement. Despite its simplicity, this potential model captures the basic properties of a multitude of classical systems and serves as an important reference point for more complex confinement situations [13].

Particles are contained in a $10 \times 10 \times 10$ cm cubic box. The system is confined by the potential given in Eq. (1), with the box size chosen to be large enough that particles, confined by the external potential, always remain within the box for the various Debye lengths examined. Particles are assumed to have constant and equal masses of $m_d = 1.74 \times 10^{-12}$ kg, equal charge of $q = 3.84 \times 10^{-16}$ C and equal radii of $r_0 = 6.5$ μm. Initially all simulations assume a random distribution of particles subject to the condition that the center of mass of the particle system must be located at the center of the box. Thermal equilibrium for the particle system at a specified temperature is established by allowing collisions of the dust particles with neutral gas particles. Collisions are assumed to be elastic with both momentum and kinetic energy conserved and $10^5$ collisions for each particle in each time step allowed. For this research, in order to investigate the structure and energy of the system in the ground and first few excited states, all simulations are conducted at a temperature $T = 0$ (i.e., the neutral gas particles are held motionless). Particles are initially given a zero velocity; they obtain higher velocities almost immediately after the start of the simulation due to potential interactions and then cool via collisions with neutrals. Crystal formation occurs less than 5 seconds into the simulation with the particle system continuing to cool slowly thereafter and reaching a temperature of around 1 degree K at approximately 30 seconds. Results are not dependent on particle radius since particles are modeled as point masses where the radius as given above is used only for calculation of particle mass. Dimensionless lengths and energies are employed throughout by introducing the units $r_0 = (q^2/2\pi\varepsilon m\omega^2)^{1/3}$ and $E_0 = (m\omega^2 q^4/32\pi^2\varepsilon^2)^{1/3}$, respectively [13].

## III. SHELL STRUCURE AND OCCUPATION NUMBERS

One primary trait of 3D unscreened Coulomb clusters is the occurrence of concentric shells with characteristic occupation numbers [2], [3], [11]-[13]. For the 3D screened Coulomb clusters investigated in this research, shell structures were also obtained, as shown in Fig.1 for clusters with $\lambda_D = 0.1389$. Fig.1 (a) shows the distance between representative particle positions and the center of the cluster for all particles in a cluster with $N = 100$ where a three-shell structure can be clearly seen while Fig.1 (b) plots the radii of these shells as a function of particle number N ($4 \leq N \leq 1400$). As particle number N increases, the cluster transitions from a one-shell system to a two-shell, three-shell or higher number shell system with the shell radius increasing as well. The shell occupation numbers, where the cluster transitions to the next higher number shell, comprise a sequence, 11, 43, 120 …. Interestingly, this sequence is different from the occupation numbers calculated by previous research groups (13, 58, 155 …) for pure Coulomb clusters, where $\lambda_D = \infty$ is assumed [13]. As such, it can be seen that in this case the occupation numbers appear dependent on the system's Debye length, in agreement with previous experimental results [10].

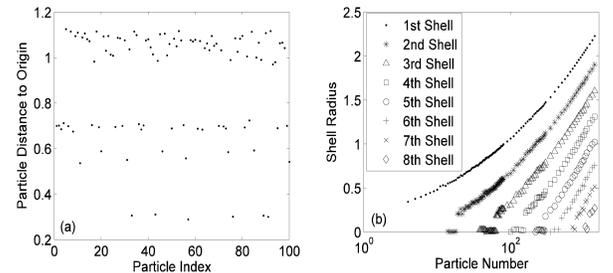

Fig. 1. (a) The distance between particle position and the center of the cluster for all particles within a cluster with $N = 100$. (b) The shell radii as a function of particle number N for $4 \leq N \leq 1400$.

To better quantify the above, simulations for clusters having various Debye lengths, were examined around the first three occupation numbers. From the resulting data, it was determined that both the occupation numbers and the radii of the shells are dependent on the Debye length. This dependence of occupation number and shell radius on Debye length for clusters having N =100 is shown in Fig.2. Fig.2 (a) shows the evolution of the second occupation number as a function of $\lambda_D$. As can be seen, for Debye lengths larger than 2-3, the function remains flat. That is, from the pure Coulomb limit (corresponding to a OCP) to 2-3, the second occupation number can be assumed independent of the Debye length. As seen in Fig.2 (b), the corresponding radii of the shells are almost constant over this range of Debye length as well. Thus, it can be concluded that cluster structure is largely independent of Debye length for clusters with $\lambda_D > 2$-3 (long Debye length). However, for $\lambda_D < 2$-3, as shown in Fig.2 (a) and (b), both the shell radii and the occupation numbers exhibit a sharp increase as the Debye length increases. Thus over this regime, cluster structure is very sensitive to any change in Debye length.

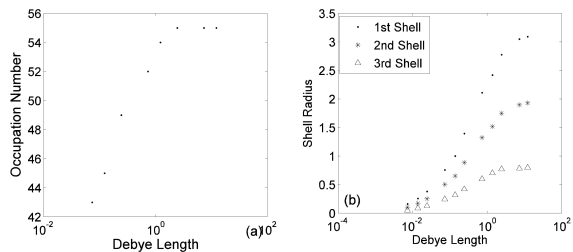

Fig. 2. (a) Second occupation number as a function of $\lambda_D$. (b) Shell radii as a function of $\lambda_D$ (The shell numbers run from outside in.)

## IV. GROUND STATES AND LOW ENERGY METASTABLE STATES

To analyze in detail the influence of screening length on the structure and energy of dust clusters, numerous simulations were conducted for particle numbers, 4<N<20. For each particle number, 6 different Debye lengths were simulated. For each Debye length, all possible ground states and low energy meta-stable states were calculated. In all, 50 simulations with different initial particle positions (provided by random number seed) were conducted. The energies and inter-shell structure of the states obtained from these simulations are shown in Table I.

The ground state energies of the clusters as functions of the Debye length are shown in Fig. 3. As can be seen in Fig. 3(a), using the 4 particle cluster as an example, the ground state energy increases as the Debye length increases. However, just like the occupation number and shell radius, this becomes almost constant for large Debye length. In comparison with previous results for unscreened Coulomb clusters, the limit of the energy as the Debye length increases to infinity is exactly the same as the ground state energy for the corresponding unscreened Coulomb cluster. Fig. 3(b) shows the ground state energies as functions of Debye length for all clusters with particle number 4<N<20. As shown, the shape of the $\lambda$-E function does not depend on particle number. Each curve, representing the dependence of the ground state energy on Debye length, shows exactly the same trend with no overlap between curves.

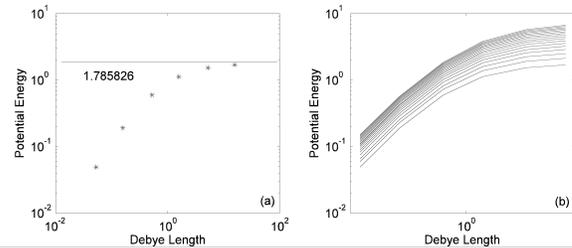

Fig. 3. (a) The ground state energy as a function of Debye length for the 4 particle cluster. The solid line shows the value of the unscreened Coulomb cluster. (b) The ground state energy as a function of Debye length for all particle numbers.

Table I shows many entirely new states for finite Debye length clusters, with structures (either inter-shell or intra-shell) never been seen for unscreened Coulomb clusters. As a representative example, only a single shell structure is obtained for 9-particle clusters in the unscreened case [13]. However, it can be seen in Table I that a two-shell meta-stable state with 8 particles on the outside shell and 1 particle at the center is obtained for finite Debye lengths. An example for new states with new intra-shell structures will be discussed in Sec. V (Fig. 4(b)). It can also be seen that as the Debye length changes, there can be corresponding structural transitions, i.e. the ground state may change from one structure type to another.

## V. INTER-SHELL AND INTRA-SHELL STRUCTURAL TRANSITIONS

Table I also shows that for certain cases, the ground state can change from one type of inter-shell structure to another as the Debye length changes. For example, for $\lambda_D \geq 1.5754$, the ground state for the 11-particle cluster is a single-shell structure but for $\lambda_D \leq 0.5251$, it becomes a two-shell state with 10 particles on the outside shell and 1 particle at the center. Thus, there is an inter-shell structural phase transition between the two states as the Debye length changes. As seen in Table I, this type of structural phase transition between 1-shell and 2-shell states also appears for 12-particle clusters. A second example of such an inter-shell structural phase transition can be seen in 19 and 20 particle clusters, where there is a transition between a 2-shell state with 1 particle at the center and a 2-shell state with 2 particles at the center.

To determine the critical value of the Debye length for these transitions, additional simulations around the transition points were conducted. As can be seen in Tables II and III, the critical values of Debye length determined in this manner for the 1-2 shell transition in the 11 and 12 particle clusters are 0.5251 and 0.2468 respectively.

Table II. Potential energy of the 1 shell and 2-shell structural states for the 11 particle cluster around the 1-2 shell transition point. (For very low energy meta-stable states, the energy difference with respect to the ground state is given in brackets.)

| λ | 0.3970 | 0.4631 | 0.5251 | 0.5955 | 0.6616 |
|---|---|---|---|---|---|
| 1-shell | 1.0087 | 1.1510 | 1.2752 | 1.4063 | 1.5210 |
| 2-shell | 1.0073 | 1.1502 | (2.1934e-6) | 1.4073 | 1.5227 |

Table III. Same as Table II but for 12 particle clusters

| λ | 0.1575 | 0.2101 | 0.2468 | 0.3151 | 0.3676 |
|---|---|---|---|---|---|
| 1-shell | 0.4198 | 0.5731 | 0.6765 | 0.8589 | 0.9905 |
| 2-shell | 0.4184 | 0.5725 | (5.4943e-6) | 0.8602 | 0.9928 |

It is important to note that the inter-shell structure characterized by the occupation numbers on each shell as shown in Table I is not sufficient to specify a state. At times, there can be more than one state having the same inter-shell structure. In such cases, intra-shell structures must be specified in order to distinguish between the states with two types of intra-shell structures prevalent. For the first of these (type I), voronoi numbers can be used to specify the intra-structure for that shell. For example, the outermost shell structure for the cluster shown in Fig. 4(a) can be noted as [0,0,0,0,5,12,0,0]. The number in the nth digit of the sequence represents the number of particles having the voronoi number n, i.e., having n neighboring particles. Thus, the cluster shown in Fig. 4(a) has five 5-neighbor particles, and twelve 6-neighbor particles on the outermost shell. For type II, the voronoi number is not adequate alone to specify the intra-shell structure since several having exactly the same voronoi number sequences can exhibit completely different symmetries. In such cases, the voronoi symmetry parameters must also be used to determine the structure [13].

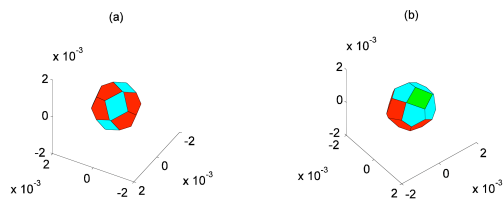

Fig. 4. The spherical voronoi diagrams for the outermost shell structure of the 18 particle cluster denoted as (a) [0,0,0,0,5,12,0,0] for $\lambda_D = 0.5$ and (b) [0,0,0,4,7,6,0,0] for $\lambda_D = 0.16$.

Representative cases exhibiting type I intra-shell structural phase transitions, i.e. where the ground state structure changes from one type to another, are shown in Table I. For example, for the 18-particle clusters, a 2-shell state with the outmost shell having structure [0,0,0,0,5,12,0,0] (Fig.4 (a)) is the ground state for $\lambda_D = 0.5$, however, for $\lambda_D = 0.16$, another 2-shell state with the outmost shell having structure [0,0,0,4,7,6,0,0] (Fig. 4 (b)) becomes the ground state. Intra-shell structural phase transitions for type II have not yet been observed. For example, as seen in Table I, although for N = 17, there are two (1, 16) states for each Debye length, both of which are type II states as discussed in [13], no phase transitions are observed. In this case, for all Debye lengths, the ground state is always the state with the higher Voronoi symmetry.

## VI. CONCLUSIONS

The crystallization of a 3D Yukawa system assuming a parabolic spatial confinement was simulated using the box_tree code. Simulations were conducted for particle numbers over the range $4 \leq N \leq 1400$ and for various values of Debye length. Shielded 3D Coulomb clusters were shown to exhibit structures comprised of concentric shells having characteristic occupation numbers. Shell radii as a function of particle number were calculated for $\lambda_D = 0.1389$ and it was determined the occupation numbers for systems having finite Debye lengths are different than those calculated for pure 3D Coulomb systems.

Simulations for clusters having various Debye lengths were also run around the first three occupation numbers. From the pure Coulomb limit (corresponding to a OCP with $\lambda_D = \infty$) to $\lambda_D = 2$-3, it was shown that both magic numbers and shell radii can be considered independent of the Debye length. However, for $\lambda_D < 2$-3, both the shell radii and the occupation numbers exhibit a sharp increase as Debye length increases, implying that cluster strucure is sensitive to any change of Debye length over this range.

The ground states and low energy meta-stable states were obtained from box_tree simulations for particle numbers $4 \leq N \leq 20$ and the ground state energies of the clusters as a function of the Debye length were analyzed. Similar to the occupation numbers and shell radii, the ground state energies increase as the Debye length increases but become almost constant for large Debye length with the energy approaching the ground state energy for an unscreened Coulomb cluster in the limit as the Debye length goes to infinity (OCP). Interestingly, a graph of the potential energies as a function of the Debye lengths for all clusters with $4 \leq N \leq 20$ shows no overlap between the $\lambda_D$-E curves; thus, the shape of the $\lambda_D$-E function does not depend on the particle number.

Cluster structure for different states was also analyzed for various Debye lengths. Many entirely new states for finite Debye length clusters, with structures (either inter-shell or intra-shell) not seen for unscreened Coulomb clusters were discovered. Again, structural phase transitions, including inter-shell structural phase transitions and intra-shell structural phase transitions, were observed and investigated for changing Debye length.

Table I. Shell configuration, potential energy for the low energy states (for the very low energy meta-stable states the energy difference with respect to the ground state is given in brackets).

|   | 15.7538 | 5.2513 | 1.5738 | 5.2513e-1 | 1.5738e-1 | 5.2513e-2 |
|---|---|---|---|---|---|---|
| 4 | (4)1.6943 | (4) 1.5317 | (4)1.1232 | (4) 0.5967 | (4)0.1904 | (4) 0.0490 |
| 5 | (5)2.1235 | (5) 1.9093 | (5)1.3819 | (5) 0.7243 | (5)0.2302 | (5) 0.0594 |
| 6 | (6)2.5024 | (6) 2.2373 | (6)1.5965 | (6) 0.8215 | (6)0.2566 | (6) 0.0655 |
| 7 | (7)2.8826 | (7) 2.5674 | (7)1.8173 | (7) 0.9302 | (7)0.2919 | (7) 0.0752 |
| 8 | (8)3.2320 | (8) 2.8674 | (8)2.0117 | (8) 1.0210 | (8)0.3197 | (8) 0.0826 |
| 9 | (9)3.5687<br>(1,8)3.6014 | (9) 3.1551<br>(1,8) 3.1877 | (9)2.1969<br>(1,8)2.2275 | (9) 1.1077<br>(1,8) 1.1300 | (9)0.3462<br>(1,8)0.3560 | (9) 0.0894<br>(1,8) 0.0929 |
| 10 | (10)3.8943<br>(1,9)3.9162 | (10) 3.4323<br>(1,9)3.4550 | (10)2.3744<br>(1,9)2.3939 | (10) 1.1915<br>(1,9) 1.2032 | (10)0.3729<br>(1,9)0.3757 | (10) 0.0968<br>(10) 0.0970<br>(1,9) 0.0973 |
| 11 | (11)4.2130<br>(1,10)4.2228 | (11) 3.7031<br>(1,10) 3.7127 | (11)2.5481<br>(1,10)2.5551 | (1,10) 1.2752<br>(11) (2.1934e-6) | (1,10)0.3962<br>(11)0.4004 | (1,10) 0.1022<br>(1,10) 0.1028<br>(11) 0.1043 |
| 12 | (12)4.5092<br>(1,11)4.5245 | (12) 3.9518<br>(1,11) 3.9668 | (12)2.7017<br>(1,11)2.7142 | (12) 1.3432<br>(1,11) 1.3482 | (1,11)0.4184<br>(12)0.4198 | (1,11) 0.1079<br>(12) 0.1092 |
| 13 | (1,12)4.8076<br>(1,12)4.8127 | (1, 12) 4.2027<br>(13) 4.2082 | (1,12)2.8569<br>(13)2.8655 | (1, 12) 1.4089<br>(13) 1.4241 | (1,12)0.4341<br>(1,12)0.4349<br>(13)0.4482 | (1, 12) 0.1112 |
| 14 | (1,13)5.0973<br>(14)5.1009 | (1,13) 4.4458<br>(14)4.4497 | (1,13)3.0090<br>(14)3.0161 | (1,13) 1.4811 | (1,13)0.4583 | (1,13) 0.1182 |
| 15 | (1,14)5.3742<br>(15)5.3833 | (1,14) 4.6763 | (1,14)3.1501 | (1,14) 1.5452 | (1,14)0.4781 | (1,14) 0.1235 |
| 16 | (1,15)5.6463<br>(16)5.6585 | (1,15) 4.9023 | (1,15)3.2885 | (1,15) 1.6090 | (1,15)0.4985 | (1,15) 0.1292 |
| 17 | (1,16)5.9123<br>(1,16)5.9126 | (1,16) 5.1226<br>(1,16) 5.1230 | (1,16)3.4226<br>(1,16)3.4230 | (1,16) 1.6705<br>(1,16) 1.6711 | (1,16)0.5182<br>(1,16)0.5189<br>(2,15)0.5224 | (1,16) 0.1346<br>(1,16) 0.1350<br>(2,15)0.1364 |
| 18 | (1,17)6.1733 | (1,17) 5.3383 | (1,17)3.5536<br>(1,17)3.5537 | (1,17) 1.7312<br>(1,17) (2.5712e-5)<br>(2,16) 1.7364 | (1,17)0.5384<br>(2,16)0.5400<br>(2,16)0.5402 | (1,17) 0.1404<br>(1,17) (4.6669e-5)<br>(2,16) 0.1410 |
| 19 | (1,18)6.4296 | (1,18) 5.5494 | (1,18)3.6813 | (2,17) 1.7929<br>(2,17) (3.2196e-5) | (2,17)0.5575<br>(1,18)0.5577<br>(1,18)0.5591 | (2,17) 0.1456<br>(2,17) 0.1457<br>(1,18)0.1457 |
| 20 | (1,19)6.6835<br>(2,18)6.6878 | (1,19) 5.7586<br>(2,18) 5.7626 | (1,19)3.8085<br>(2,18)3.8105 | (2,18) 1.8482<br>(1,19) 1.8507 | (2,18)0.5749<br>(3,17)0.5772<br>(1,19)0.5785 | (2,18) 0.1504<br>(2,18) 0.1506 |

Table II. Potential energy of the 1 shell and 2-shell structural states for the 11 particle cluster around the 1-2 shell transition point. (For very low energy meta-stable states, the energy difference with respect to the ground state is given in brackets.)

| λ | 0.3970 | 0.4631 | 0.5251 | 0.5955 | 0.6616 |
|---|---|---|---|---|---|
| 1-shell | 1.0087 | 1.1510 | 1.2752 | 1.4063 | 1.5210 |
| 2-shell | 1.0073 | 1.1502 | (2.1934e-6) | 1.4073 | 1.5227 |

Table III. Same as Table II but for 12 particle clusters

| λ | 0.1575 | 0.2101 | 0.2468 | 0.3151 | 0.3676 |
|---|---|---|---|---|---|
| 1-shell | 0.4198 | 0.5731 | 0.6765 | 0.8589 | 0.9905 |
| 2-shell | 0.4184 | 0.5725 | (5.4943e-6) | 0.8602 | 0.9928 |

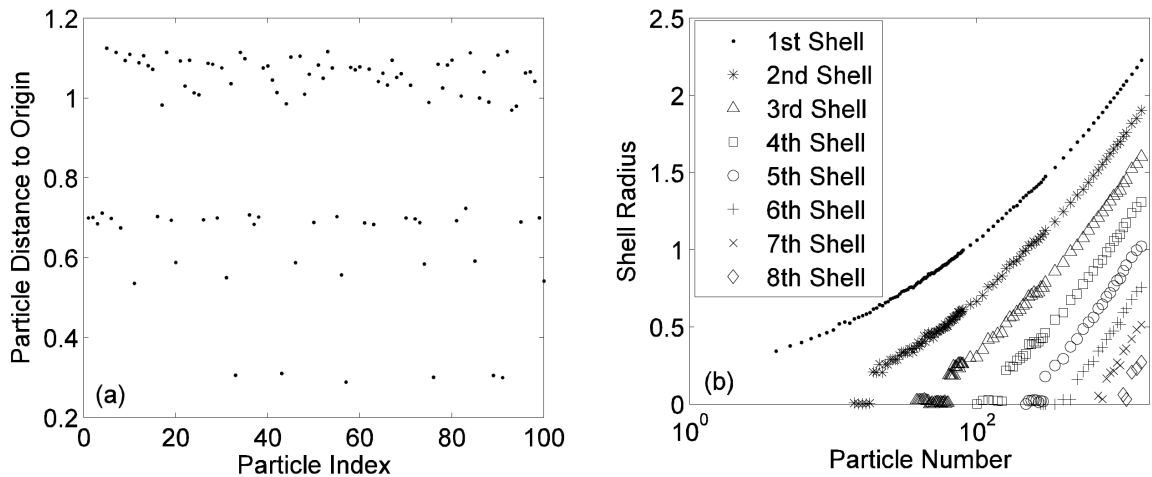

Fig. 1. (a) The distance between particle position and the center of the cluster for all particles within a cluster with N = 100. (b) The shell radii as a function of particle number N for $4 \leq N \leq 1400$.

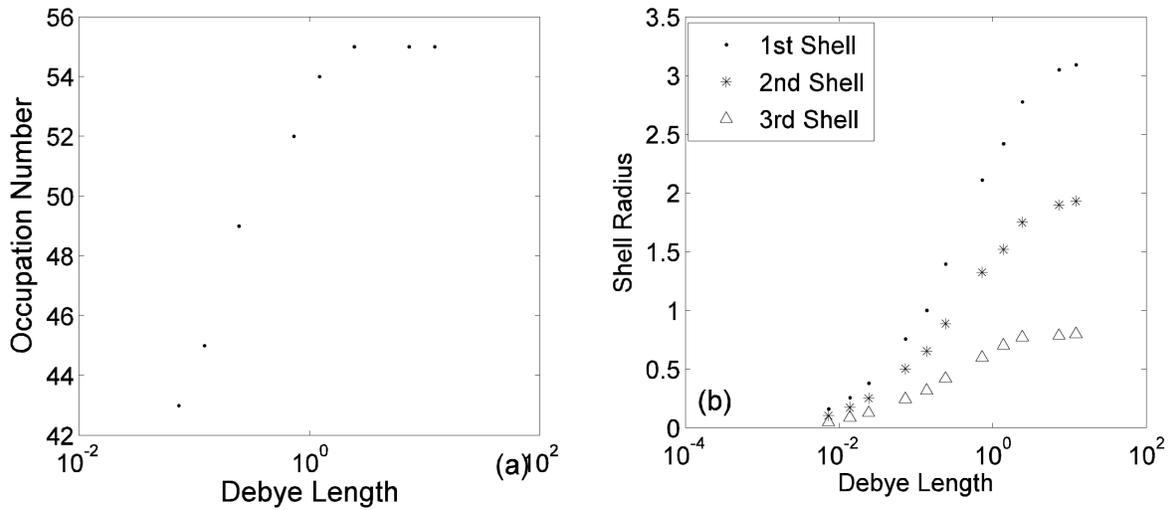

Fig. 2. (a) Second occupation number as a function of $\lambda_D$. (b) Shell radii as a function of $\lambda_D$ (The shell numbers run from outside in.)

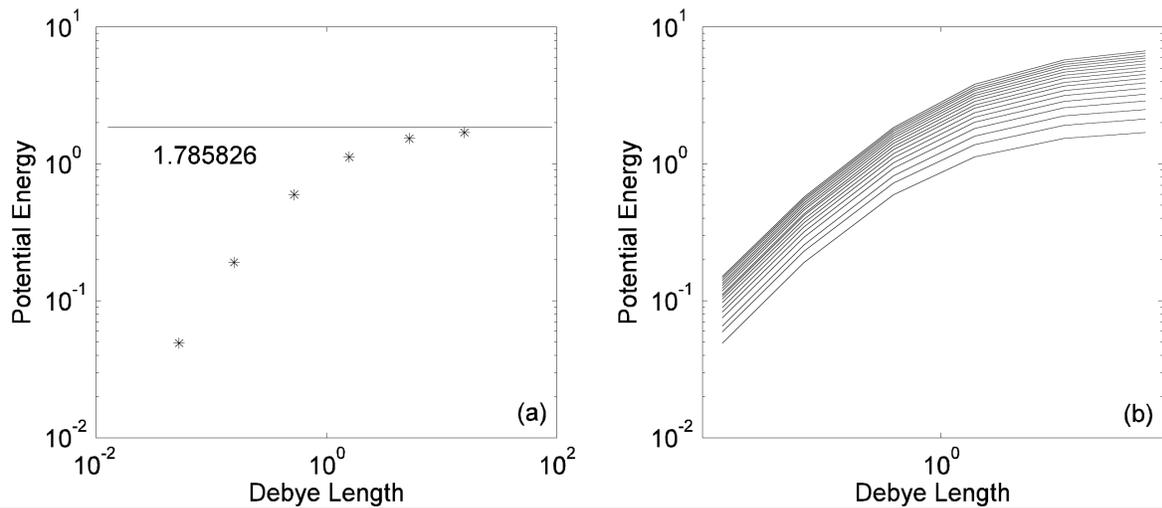

Fig. 3. (a) The ground state energy as a function of Debye length for the 4 particle cluster. The solid line shows the ground state energy for an unscreened Coulomb cluster. (b) The ground state energy as a function of Debye length for all particle numbers.

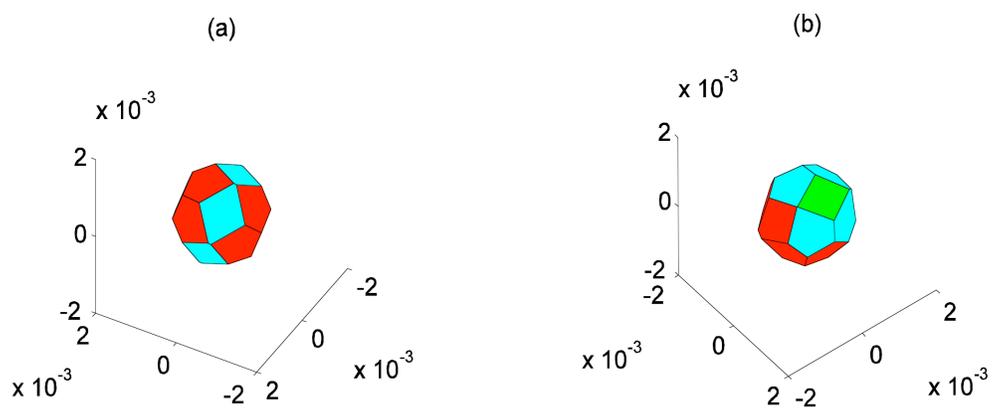

Fig. 4. The spherical voronoi diagrams for the outermost shell structure of the 18 particle cluster noted as (a) [0,0,0,0,5,12,0,0] and (b) [0,0,0,4,7,6,0,0].